\documentclass{aipproc}
\layoutstyle{6x9}
\usepackage{graphicx}
\def\sax{{\it Beppo}SAX$~$}
\classification{}
\begin{document}

\title{Spectral changes during six years of Scorpius X-1 monitoring with BeppoSAX Wide Field Cameras }

\author{Patrizia Santolamazza}{
  address={ASI Science Data Center, c/o ESA-ESRIN, via Galileo Galilei, I-00044 Frascati, Italy}
}
          \author{Fabrizio Fiore}{
  address={INAF- Osservatorio Astronomico di Roma, via di Frascati 33, I-00040 Monteporzio, Italy}
}
          \author{Luciano Burderi}{
  address={INAF- Osservatorio Astronomico di Roma, via di Frascati 33, I-00040 Monteporzio, Italy}
}
          \author{Tiziana Di Salvo}{
  address={ Dipartimento di Scienze Fisiche ed Astronomiche, Universita' di Palermo, via Archirafi 36, 90123 Palermo, Italy}
}

\begin{abstract}
We analyse a sample of fifty-five observations of Scorpius X-1 available in the \sax Wide Field Camera  public archive and spanning over the six years of \sax mission life.
Spectral changes are initially analysed by inspection of  colour-colour and colour-intensity diagrams, we also discuss the shift of the Z tracks in these diagrams. Then we select two long observations for spectral fitting analysis, a secular shift is evident between the tracks in these observations. 
We finally extract spectra along the tracks and discuss the best fit model, the parameter variations along the track and between tracks, and their link to the accretion rate.
\end{abstract}

\maketitle

\section{Introduction}
Neutron star low mass X-ray binaries (LMXB) are divided into two classes based upon the different morphology of their tracks in colour-colour and colour-intensity diagrams and their correlated timing behaviour. Six of the brightest LMXBs are classified as Z-type since they trace a Z-shaped pattern on such diagrams. The three branches of the Z are named  Horizontal (HB), Normal (NB) and Flaring (FB) branches respectively, while  atoll-type sources are characterized by an ``island'' and a ``banana'' state \cite{Has89}. The spectral state of LMXBs is most easily determined by using X-ray colour-colour and colour-intensity diagrams, in fact track morphology is due to spectral variations on timescales of weeks, days or hours. Tracks are commonly interpreted as an accretion sequence \cite{Hasetal89}, since it is thought that the mass accretion rate $\dot M$ is the parameter governing spectral variations along the track (see however \cite{bcbc03}). 
Secular shifts and shape changes of the Z track  were reported for several Z sources such as CygX-2 \cite{Kuul96}, whose shifts were interpreted in terms of occultation of the emitting region by a precessing accretion disk,  or very recently for LMC X-2 \cite{shk03}, in which secular shifts of 2.5-10\% are seen. Conversely significant secular variations of the track of Scorpius X-1 were never observed \cite{dvdk00}.
Scorpius X-1 is the brightest extra-solar X-ray source and is a LMXB of the Z-type showing a high level of activity in the X-ray, optical and radio bands, where radio jets were recently observed. In the case of Scorpius X-1, the complete Z track is generally traced out in a few hours to a day. 
 Its timing and spectral properties were studied by using data from many observatories such as EXOSAT \citep{dvdk00} and RXTE \citep{bcbc03, BrGeFo03}. This is the first systematic study of Scorpius X-1 using BeppoSAX Wide Field Cameras (WFC). 
WFCs are two coded mask instruments (WFC1 \& WFC2) with a wide field of view of 40x40 deg, pointing away from each other and perpendicular to the Narrow Field Instruments. 
In the large majority of cases WFCs observed random sky positions during primary NFI observations, giving the opportunity to monitor a large number of sources over the full six year satellite lifetime. Here we select fifty-five observations for a total monitoring duration of more than 600 hr and a total net exposure of the source of about 200 hr. 

\section{Colour-Colour and Colour-Intensity Diagrams}
In order to obtain colour-colour and colour-intensity diagrams we define the total intensity as the count rate in the 1.7-19.1 keV band, the soft colour as the ratio [3.5-6.4keV / 2.0-3.5keV] and the hard colour as the ratio [9.5-16.4keV / 6.4-9.5keV]. 

\begin{figure}[h]
  \includegraphics[height=.27\textheight, angle=-90]{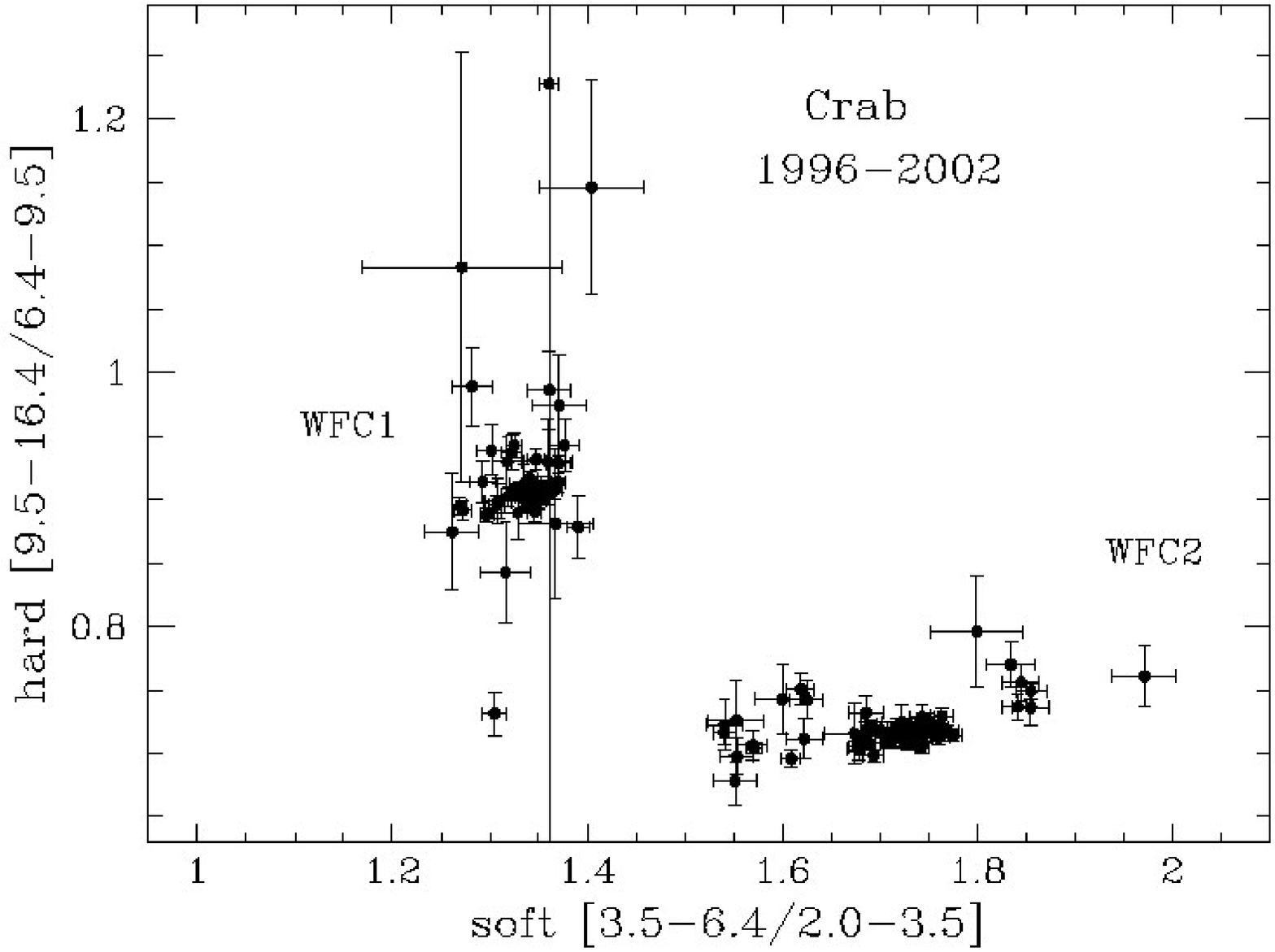}
\hspace{1cm}
  \includegraphics[height=.27\textheight, angle=-90]{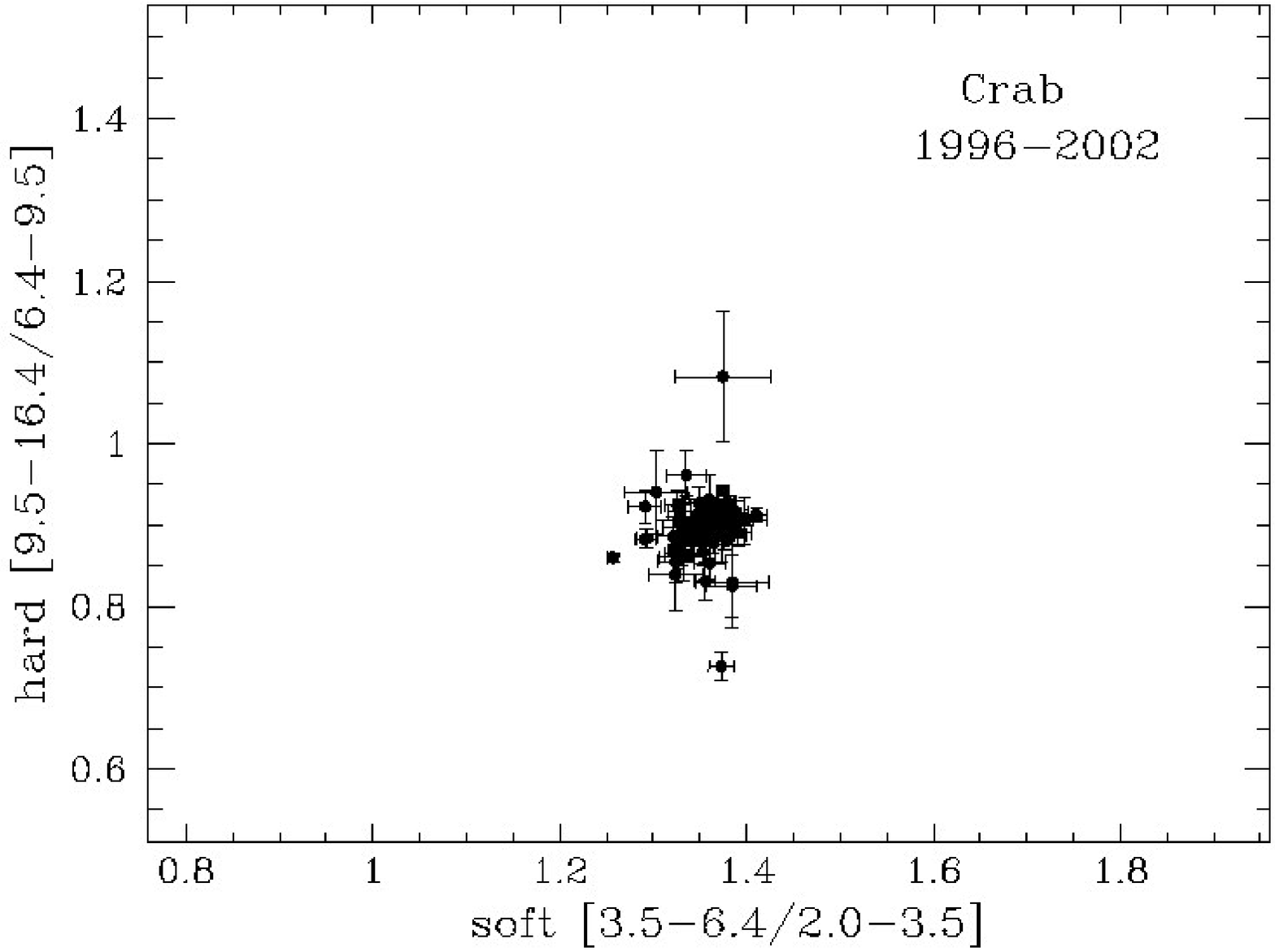}
  \caption{Crab colour-colour diagram over six years, each point represents an observation. Left panel shows raw data, the separation between WFC1 (upper left clump of points) and WFC2 (lower right clump) observations is apparent. Right panel shows offset-corrected scaled data, the diagram dimensions are on the same scale of Scorpius X-1 colour-colour diagram in right panel of Figure \ref{ccScoraw}  for comparison.}
\label{ccCrabraw}
\end{figure}
\begin{figure}[h]
  \includegraphics[height=.27\textheight, angle=-90]{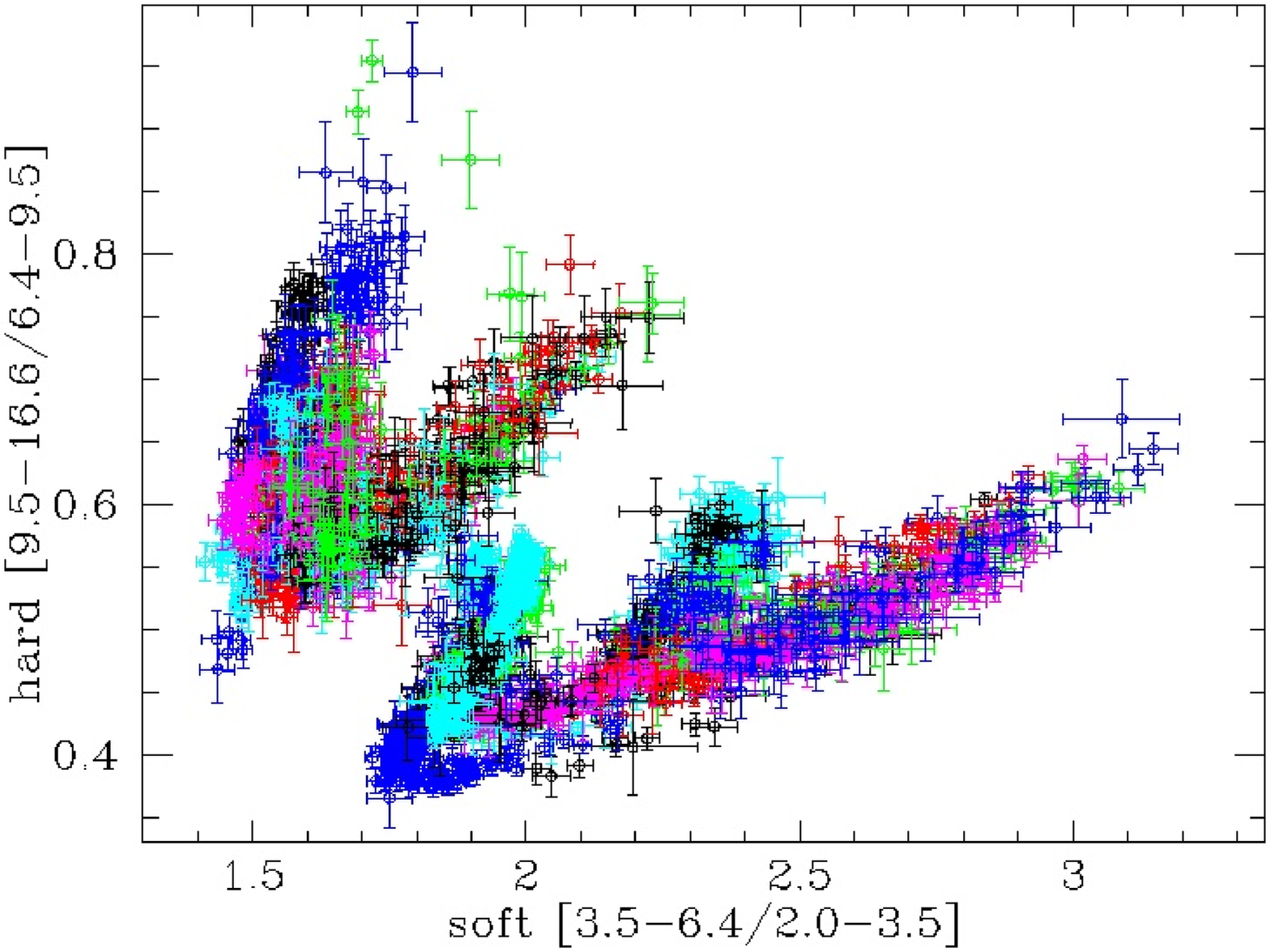}
\hspace{1cm}
  \includegraphics[height=.27\textheight, angle=-90]{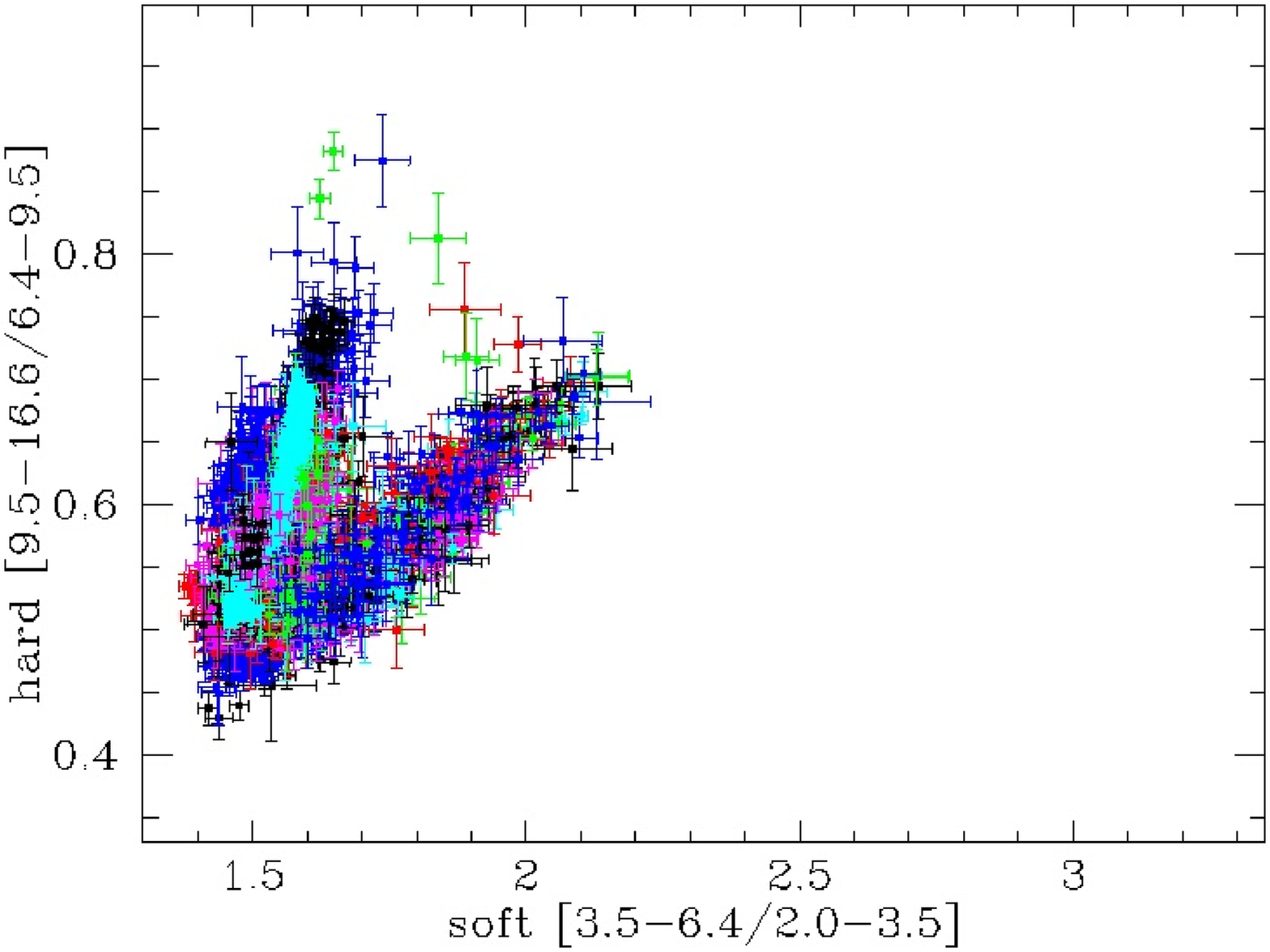}
  \caption{Scorpius X-1 colour-colour diagram over six years. Left panel shows raw data, the separation between WFC1 (upper left clump of tracks) and WFC2 (lower right clump) observations is apparent. Right panel shows offset-corrected scaled data. }
\label{ccScoraw}
\end{figure}
Since we want to compare, on these diagrams, observations pointed off the source by different offset angles and at different epochs, the detector aging and the spatial response variations must be taken into account by scaling data to a common reference condition, for instance to the center of the detector at a certain epoch. 
Scaling factors were calculated taking into account the response of each camera at the source position in the field of view (FOV) for each epoch, then  count rates and hardness ratios were scaled to the central quadrant of WFC1 at the date January 2002. 
These scaling factors change the intensity by 1\% at most, while for the soft colour and the hard colour the maximum change is 15\% and 8\% respectively.
Systematic residuals were checked using observations of the Crab Nebula and an empirical correction to systematic residual effects was estimated \citep{pat1} as a function of the off-axis angle. 
The maximum empirical corrections found are 13\%, 7\%, 2\% for the intensity, hard colour and soft colour respectively, over the 0 to 20 deg range. Therefore corrections are applied to intensity and hard colour only.
Figure \ref{ccCrabraw} shows the Crab Nebula colour-colour diagram. A clear separation between WFC1 and WFC2 data and a noticeable spread within each camera data set is seen in the uncorrected diagram (left panel), while  points from both WFCs overlap in the corrected diagram (right panel). Average Crab colours are: <soft>=1.34, <hard>=0.89 with a spread of 3$\sigma$/<soft>=5\%  3$\sigma$/<hard>=7\%.
\begin{figure}[h]
  \includegraphics[scale=.3]{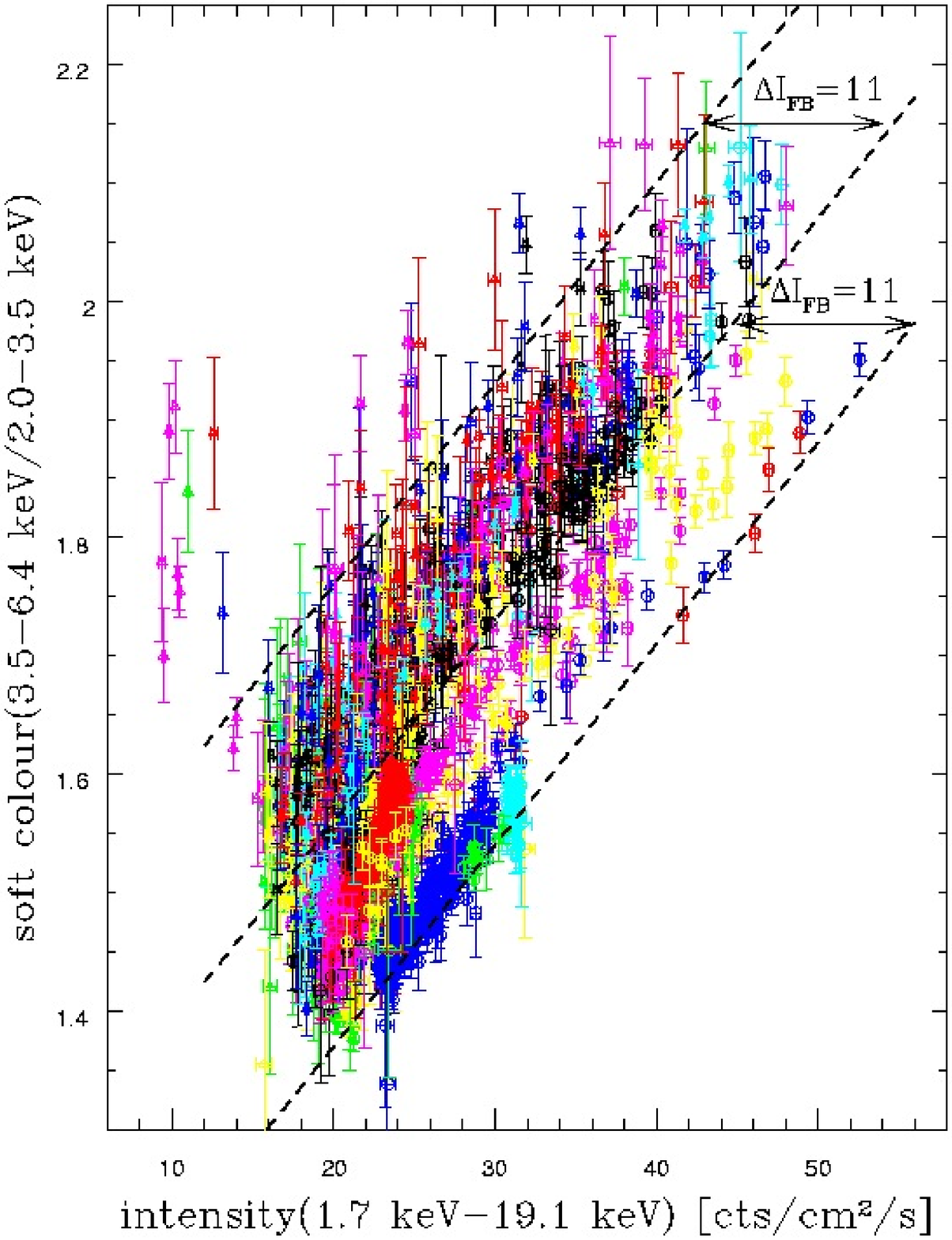}
\hspace{1cm}
  \includegraphics[scale=.3]{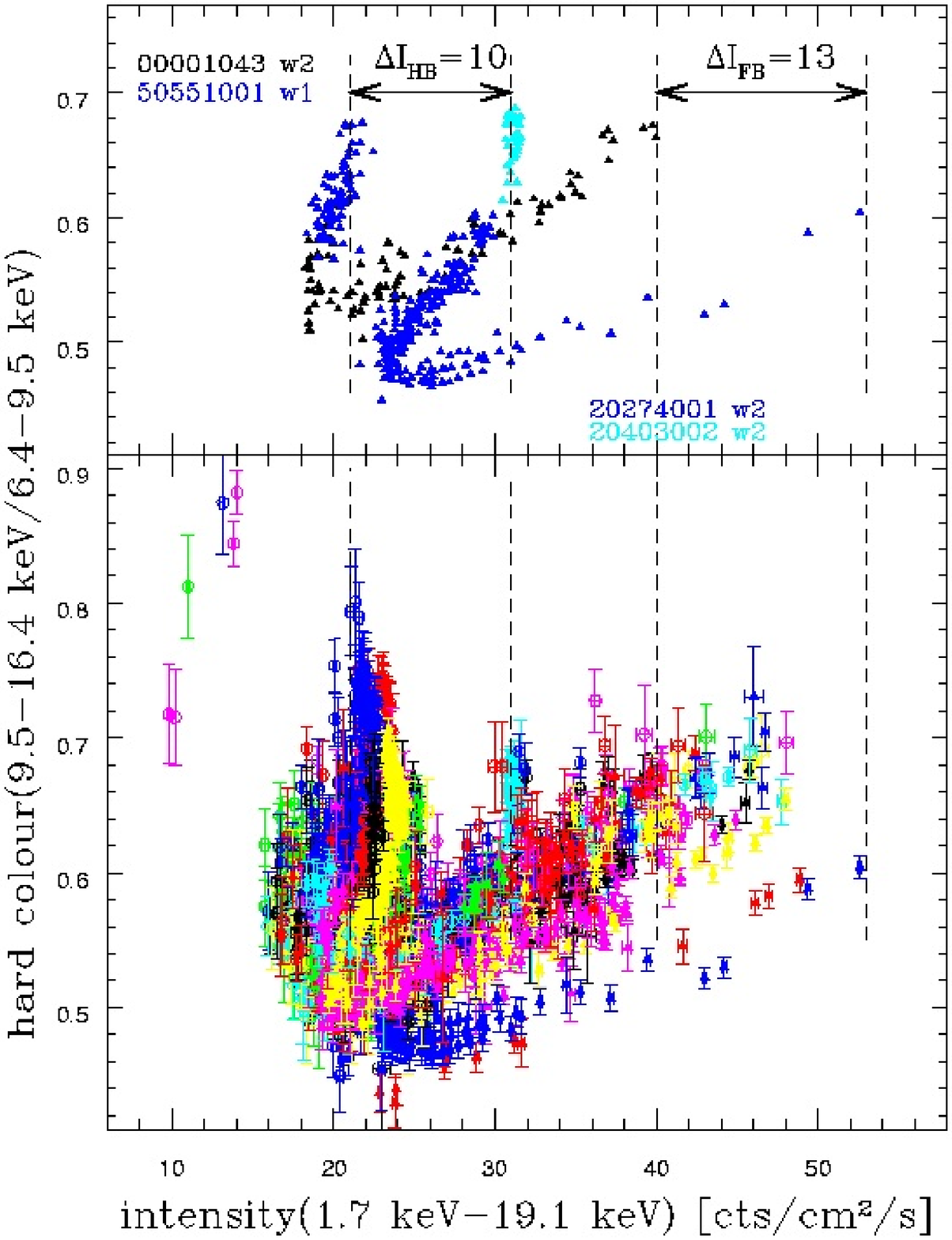}
  \caption{Scorpius X-1 corrected colour-intensity diagrams over six years.  Parallel tracks appear due to large intensity variations. Arrows indicate line spacing of $\Delta$I = 10-13 cts/cm$^2$/s. Left panel shows the soft colour versus intensity, the dashed lines mark, from left to right, the upper envelope to all data, the lower envelope to the bulk of ``normal'' observations and the lower envelope to all data. The right panels show the hard colour versus intensity, the lower panel includes all observations while the upper panel shows only two normal and two shifted observations. The vertical dashed lines mark, from left to right, the two most widely intensity-shifted (almost vertical) horizontal branches and the tip of the  the two most widely intensity-shifted flaring branches. }
\label{ci}
\end{figure}

Figure \ref{ccScoraw} shows Scorpius X-1 colour-colour diagrams over six years. In the left panel a clear separation appears between the Z-tracks measured with WFC1 and located in the upper left part of the diagram and those measured with WFC2 and located in the lower right part. The right panel shows scaled and offset-corrected data, a spread is still present in this diagram, vertex points can differ by as much as $\Delta$soft=6\% and $\Delta$hard=14\%. Thus observations suggest that secular variations are present. We checked that about 10\% of tracks in the sample is shifted with respect to the remainder of ``normal'' observations.
Secular shifts are more evident in the hardness-intensity diagrams in Figure \ref{ci}, where some tracks are characterized by softer hardness ratios and higher total intensity. Large intensity variations produce parallel tracks in both diagrams. The soft colour is well correlated with intensity (see left panel in Figure \ref{ci}) but tracks, which look like slanting lines, may differ by up to 30\% in intensity at equal soft colour. Also in the right panels of  Figure \ref{ci} the tips of the flaring branches may differ by 30\% and horizontal branches by 50\% in intensity, while in the Crab hardness-intensity diagram (not shown) the largest percent variation for intensities is 16\%. 
It is known that on long timescales the total luminosity of LMXBs may vary secularly and that temporal variability indices like the QPO frequency 
and spectral parameters form a set of parallel lines when plotted versus the total luminosity, each line reflecting the short-term correlation with luminosity and the offset reflecting the average luminosity difference in different epochs (see \cite{MvKF01, vanderkli01}). A possible explanation for the ``parallel tracks'' phenomenon was proposed by \cite{vanderkli01}. 

\section{Spectra}

\begin{figure}[bh]
  \includegraphics*[height=.28\textheight]{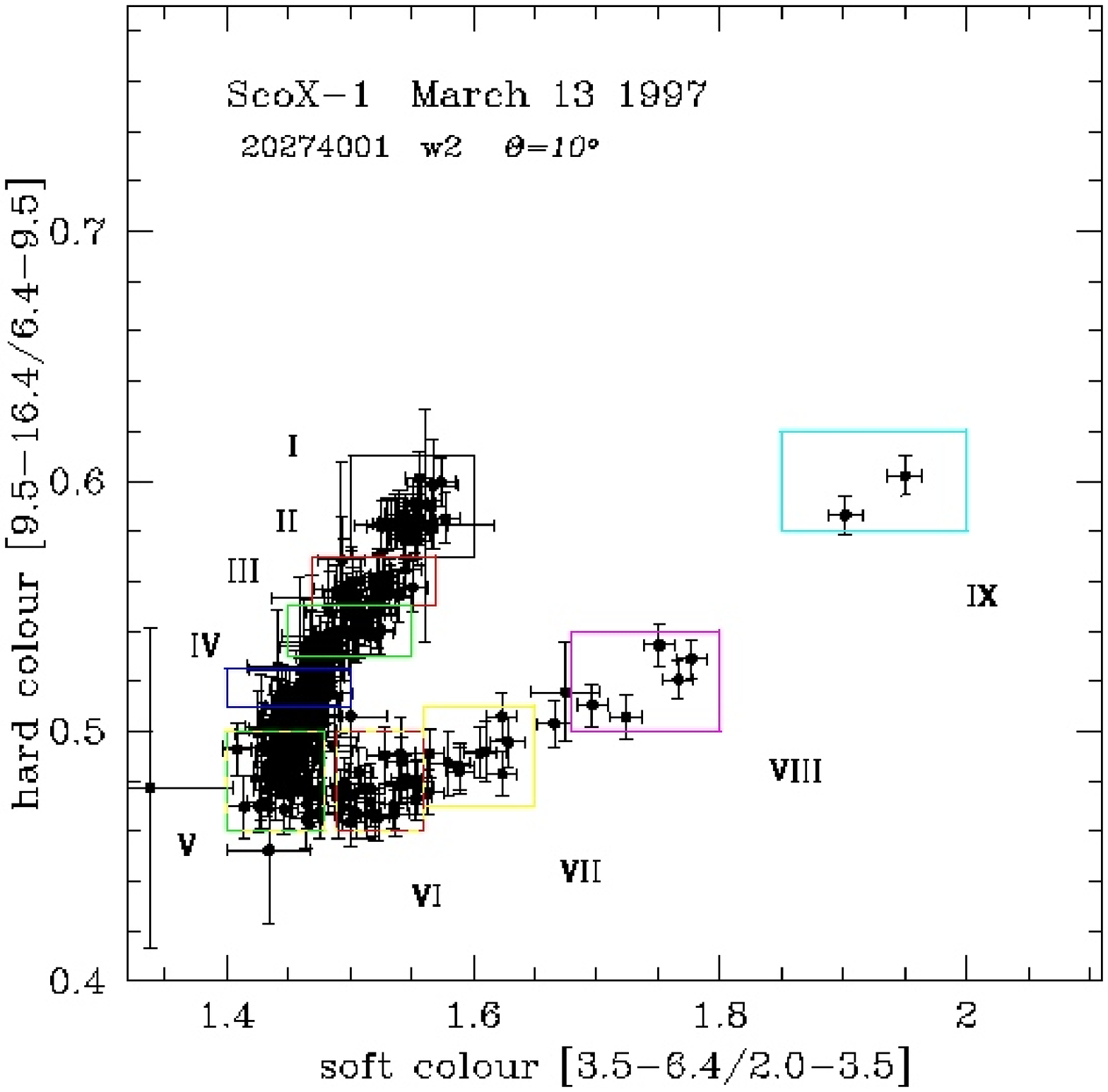}
\hspace{1cm}
  \includegraphics*[height=.28\textheight]{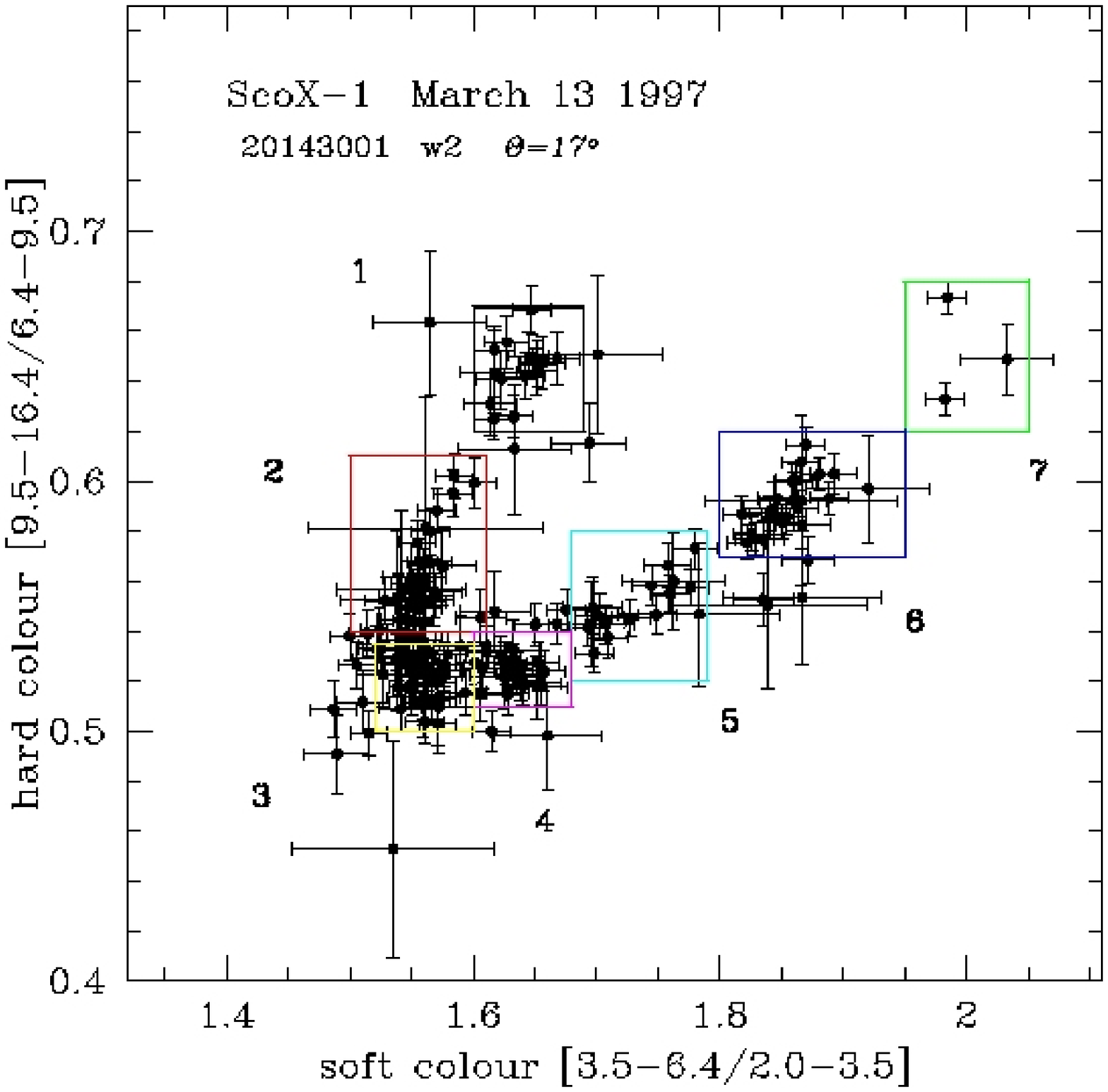}
  \caption{Scorpius X-1 corrected colour-colour diagrams of observations 20274001 (left panel) and 20143001 (right panel). The overlaid boxes delimit the data used to make spectra and are numbered from HB-NB to FB.}
\label{boxes}
\end{figure}

\begin{figure}[ht]
  \includegraphics*[height=.35\textheight]{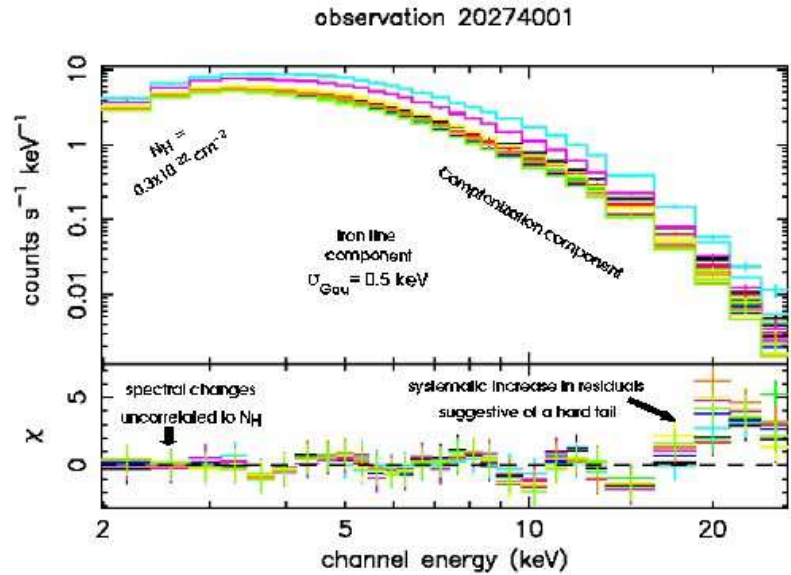}
  \caption{Spectral fits to spectra along the track of observation 20274001. Upper panel shows data and fitted models, labels indicating model components are positioned in the corresponding component energy ranges. Lower panel shows residuals in terms of sigmas, arrows indicate where expected (or unexpected!) variations occur. }
\label{spec}
\end{figure}

Spectral analysis is performed in the energy range 2.0-22 keV  along the Z-track of two observations, the ``normal'' observation 20143001 of January 25 1997 and the ``most shifted'' observation 20274001 of March 13 1997. 
Figure \ref{boxes} shows corrected colour-colour diagrams of both observations with boxes delimiting the data used to make spectra.
Our best fit model is made up of a Comptonization component, a Gaussian component with a fixed width $\sigma$=0.5 keV to model an iron line component and an absorbing column density set to the Galactic value N$_{H}$ = 0.3$\times$10$^{22}$ cm$^{-2}$. A 2\% systematic error was assumed for all spectra. 
The general behaviour of spectral changes is the same along both tracks. As an example spectra of observation 20274001 and best fit models are shown in the upper panel of Figure \ref{spec} while lower panel shows residuals in terms of sigmas with error bars of size one. Residuals are close to zero 
in the low energy range of either normal or "shifted" observations, consequently our choice of the Galactic value for N$_{H}$ seems not to affect results. Residuals at energy higher than 18 keV suggest that a further hard tail component is needed.
\begin{figure}[b]
  \includegraphics*[height=.24\textheight]{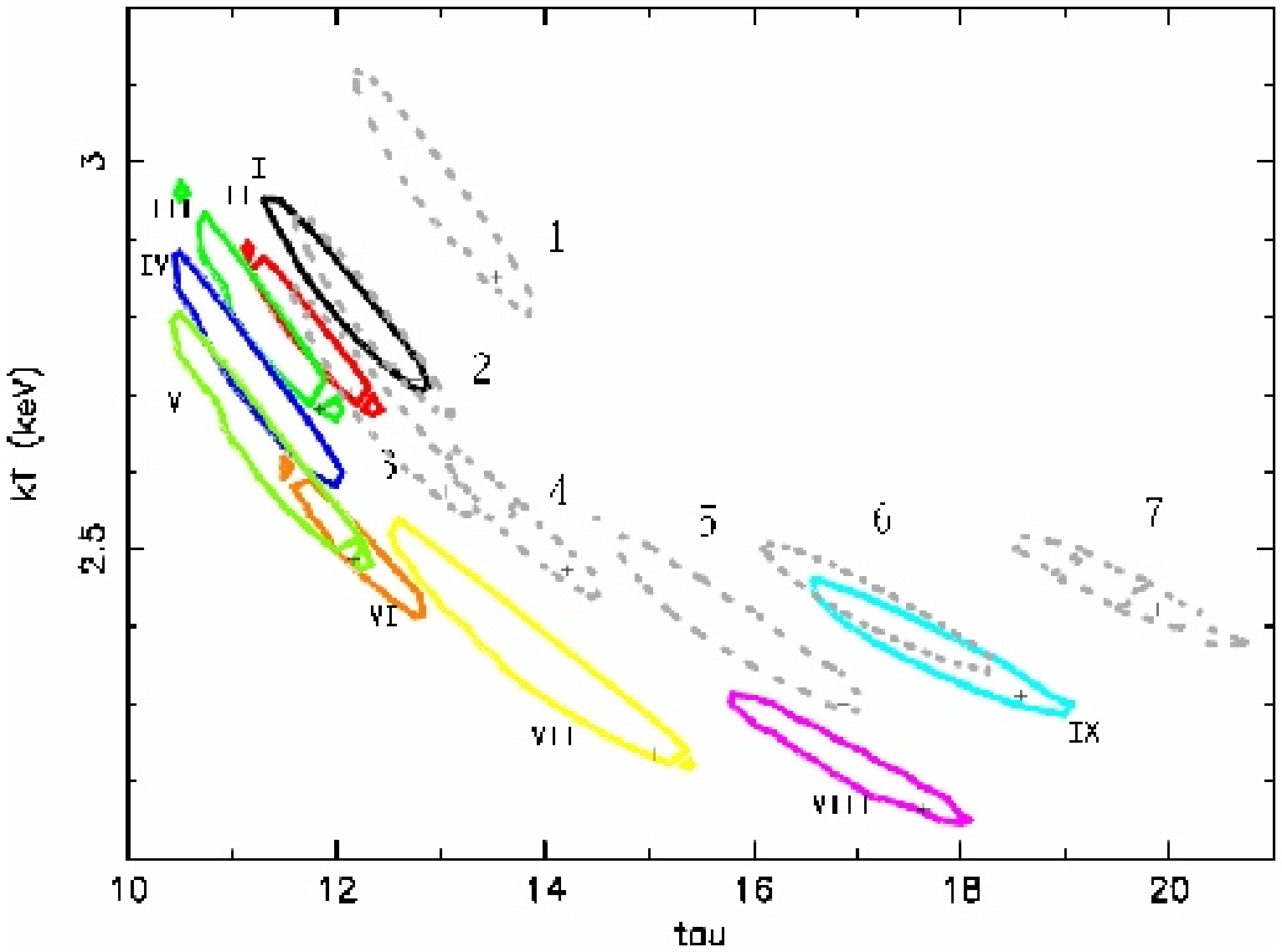}
\hspace{.5cm}
  \includegraphics*[height=.24\textheight]{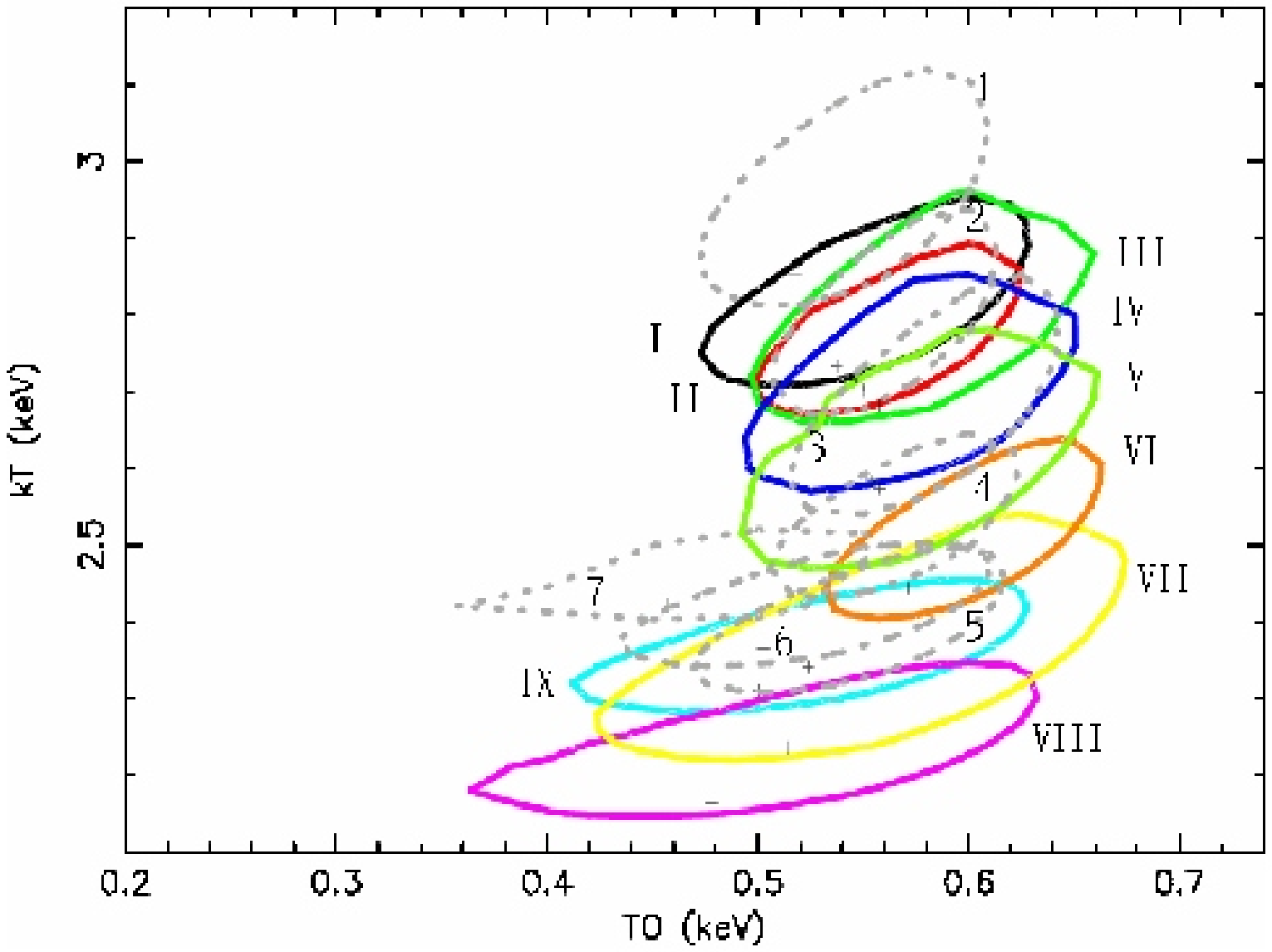}
  \caption{90\% $\chi^2$ confidence level contour plot for the Comptonization component parameters. The hot plasma temperature (kT) is reported as a function of the optical depth (tau) in left panel and of the seed photon temperature (T0) in right panel. Contour numbers correspond to spectrum boxes in Figure \ref{boxes}, solid lines belong to observation 20274001 while gray dotted lines belong to observation 20143001. }
\label{contours}
\end{figure}
 This is also supported by the comparison with Crab Nebula spectra (not shown), which show no such systematic distortion and the ratios of data over model show random distribution around unity.

In Figure \ref{contours} we report, for both observations, the 90\% $\chi^2$ confidence level contour plots for two Comptonization component parameters. The hot plasma temperature is plotted as a function of the optical depth in left panel and of the seed photon temperature in right panel. There is a strong correlation between the plasma temperature and the optical depth, the temperature decreases by about 20\% from the HB-NB to the FB while the optical depth increases by 50\% from the vertex (V), namely the transition point between NB and FB, to the FB. 
We interpret the plasma temperature decrease from HB-NB to FB in terms of Compton cooling due to increasing mass accretion rate, and therefore to increasing seed photon number. The $\dot M$ increase is accompanied by a growing number of plasma particles in the corona which contributes to the optical depth. Luminosity variations are also interpreted as an effect of the $\dot M$ increase, however we note that the total luminosity follows the intensity behaviour seen in colour-intensity plots, where the total count rate increases on the FB but slightly decreases from the HB-NB to the V. The optical depth behaves like the total luminosity.  

The luminosity along the track of the shifted observation is systematically higher by 10-20\% but also the ranges of variation of the optical depth and the plasma temperature seem to be affected. Comparison between these two parameters on corresponding branches (cfr Figure \ref{contours}) shows that their ranges move to lower values by as large as 10\%. This decrease matches with the softer hardness ratios found in the previous paragraph. So the two parameters seem to correlate to luminosity changes as average luminosity varies in different epochs.

\section{Conclusions}

Comparison of Scorpius X-1 colour-colour and hardness-intensity diagrams of fifty-five observations shows secular shifts of the tracks in 10\% of the sample. 
Spectra from two observations are fitted with a constant column density, a Comptonization component and an iron line component. 
The spectral analysis shows great spectral changes from the HB-NB to the FB. The plasma temperature decreases along the track while the optical depth increases together with the total luminosity.
The interpretation of these changes in terms of inverse Compton cooling confirms the widespread idea that it is  the mass accretion rate $\dot M$ that varies along the track.
The analysed observations show a systematic luminosity difference and shifted tracks. Comparison between spectral parameters on corresponding branches suggests that the secular shift is also accompanied by correlated secular changes in the optical depth and plasma temperature.


\begin{thebibliography}{00}
\bibitem{Has89} G. Hasinger,M. van der Klis, 1989, A\&A, 225, 79
\bibitem{Hasetal89} G. Hasinger, W. C. Priehorsky, J. Middleditch, 1989, ApJ, 337, 843
\bibitem{bcbc03} R. Barnard, M. J. Church,  M. Balucinska-Church, 2003, A\&A, 405, 237
\bibitem{Kuul96} E. Kuulkers, M. van der Klis,  B. A. Vaughan, 1996, A\&A, 311, 197
\bibitem{shk03} A. P. Smale, J. Homan, E. Kuulkers, 2003, ApJ, 590, 1035
\bibitem{dvdk00} S. W. Dieters, M. van der Klis, 2000, MNRAS, 311, 201
\bibitem{BrGeFo03} C. Bradshaw, B. J. Geldzahler,  E. B. Fomalont, 2003,  ApJ, 592, 486
\bibitem{pat1} P. Santolamazza, F. Fiore, L. Burderi, T. Di Salvo, 2003, Proc. of the BeppoSAX Symposium "The Restless High-Energy Universe", pag.644
\bibitem{MvKF01} M. Mendez, M. van der Klis,  E. C. Ford, 2001, ApJ, 561, 1016
\bibitem{vanderkli01} M. van der Klis, 2001, ApJ,561, 943
\end{thebibliography}
\end{document}